\documentclass[aps,twocolumn,superscriptaddress]{revtex4}
\usepackage{graphicx}  % needed for figures
\usepackage{dcolumn}   % needed for some tables
\usepackage{bm}        % for math
\usepackage{amssymb}   % for math
\usepackage{amsmath}
\usepackage{amsfonts}
\usepackage[colorlinks=true,linkcolor=red, citecolor=magenta]{hyperref}
\usepackage{tikz}
\usetikzlibrary{shapes,arrows}
\usepackage{siunitx}
\usepackage{physics}
\usepackage{hhline} 
\usepackage{booktabs}
\begin{document}

\title{Search for possible signals of space-time non-commutativity from ACT DR6}

\author{Vishnu Rajagopal}
\email{vishnu@hunnu.edu.cn}
\affiliation{Department of Physics and Synergetic Innovation\\
Center for Quantum Effects and Applications,\\
Hunan Normal University, Changsha, Hunan 410081, China}
\affiliation{Institute of Interdisciplinary Studies,\\
Hunan Normal University, Changsha, Hunan 410081, China}
\author{Puxun Wu}
\email{pxwu@hunnu.edu.cn}
\affiliation{Department of Physics and Synergetic Innovation\\
Center for Quantum Effects and Applications,\\
Hunan Normal University, Changsha, Hunan 410081, China}
\affiliation{Institute of Interdisciplinary Studies,\\
Hunan Normal University, Changsha, Hunan 410081, China}

\begin{abstract}

The non-commutative geometry offers an effective framework for describing physics at the Planck scale, incorporating generic quantum-gravitational effects through an intrinsic minimal length and the $\kappa$-deformed space-time stands out as a particularly well-developed model based on a Lie-algebraic type non-commutative space-time structure. We investigate the inflationary paradigm and the associated primordial perturbations within the framework of $\kappa$-deformed non-commutative space-time. By constructing a scalar field theory invariant under the $\kappa$-Poincaré algebra, we derive the deformed oscillator algebra for the field modes of the primordial perturbations resulting in the non-trivial scale dependent modification of both the scalar and tensor power spectra. Further, we compute the corresponding $\kappa$-deformed corrections to the scalar spectral index and its running, finding that the $\kappa$-deformation naturally provides a higher value for spectral index which offers a potential resolution to the discrepancy between the values reported by PLANCK and ACT. Finally we perform a comprehensive Bayesian analysis using the latest ACT DR6 data, constraining the $\kappa$-deformed non-commutative length scale to $\lambda=2.17^{+2.33}_{-1.53}\times 10^{-30}m$ at the $1\sigma$ confidence level, suggesting the precision cosmology as a powerful probe of quantum gravity phenomenology. 

\end{abstract}

\maketitle

\section{Introduction}

The unification of general relativity and quantum theory into a self consistent quantum gravity theory remains one of the most profound challenges in fundamental physics. If we accept that gravity can be quantized, the effect of quantum gravity should become evident at or near the Planck scale, which is regarded as  the minimum measurable scale \cite{min-qg}. Among the various approaches to quantum gravity, non-commutative (NC) geometry offers \cite{connes} a compelling pathway by considering a non-commutative space-time structure. This paradigm naturally incorporates a minimal length scale, beyond which the space-time losses its operational meaning, capturing the quantum gravity effects \cite{dop}. Eventually different non-commutative space-time structures and relevant physical models have been constructed.

The $\kappa$-deformed space-time \cite{kappa-review} is one such particularly well-studied NC space-time whose space-time coordinates obey the following commutation relations
\begin{equation} \label{lie}
 [\hat{x}_i, \hat{x}_j]=0,~~[\hat{x}_0,\hat{x}_i]=i\kappa^{-1}\hat{x}_i
\end{equation}
with $1/\kappa$ being the fundamental minimal length scale. These NC space-time structures replaces the standard symmetry algebra with deformed Poincare algebra. For the $\kappa$-deformed space-time this symmetry algebra is defined by the $\kappa$-Poincare algebra \cite{kpa0,kpa,kpa1,kpa2} and is shown to be closely associated with the doubly special relativity (DSR) \cite{dsr} which treats $\kappa$ as an observer independent quantity. Moreover this $\kappa$-Poincare algebra is also shown to appear as the symmetry algebra of the resulting background space-time in the low-energy limit of loop quantum gravity as well \cite{loop-kappa}. The $\kappa$-Poincare algebra has been constructed in different basis and the associated Casimir operator gives rise to different modified dispersion relations. However the physical predictions obtained from different bases of the $\kappa$-Poincare theory have been shown to be independent of the choice of basis~\cite{non-linear}. These modified dispersion relations and the underlying deformed Poincare algebra significantly affect the dynamics of quantum fields propagating in such a quantized space-times. Various aspects of the $\kappa$-deformed scalar field have been studied using the notion of $\kappa$-Poincare algebra \cite{kappa-field,kappa-field1,kappa-field2,kappa-field3}. Such scalar fields originated during the Planck epoch would possibly carry the imprints of the quantum gravity signals. Therefore it is compelling to study the scalar field driven inflation in the non-commutative space-times as it provides a crucial testing scenario to probe the quantum gravity signals and would also reveal the underlying quantum nature of space-time structures.

%The inflationary paradigm is a powerful mechanism to understand the physics of very early universe. The initial seed for the inflation is originated from the quantum fluctuations of inflaton field which eventually freezes out as classical perturbations once the modes of the inflaton fluctuations crosses the horizon and they are directly related with curvature perturbations \cite{dodelson}. These primordial perturbations later leads to the formation of all cosmic structures and are imprinted directly onto the Cosmic Microwave Background (CMB) \cite{dodelson}, which we observe today. Such inflaton fields originated during the Planck epoch would possibly carry the imprints of the quantum gravity signals. Therefore it is compelling to study the inflation in the non-commutative space-times as it provides a crucial testing scenario to probe the quantum gravity signals and would also reveal the underlying quantum nature of space-time structures.

Certain aspects of the inflation have been studied extensively in the Moyal space-time, which is a canonical type non-commutative space-time that appears in the low energy limit of string theory \cite{string}. The space-time non-commutativity, introduced using Moyal star product formalism, has been shown to induce non-trivial effects on the primordial perturbations resulting in the scale-dependent modifications of primordial power spectra and affecting the temperature fluctuations of CMB \cite{moyal-infl,moyal-infl1,moyal-infl2,moyal-infl3,moyal-infl4,moyal-infl5,moyal-infl6,moyal-infl7,moyal-infl8}. Such non-commutative modifications to the power spectra also arises while studying inflation within the context of stringy space-time uncertainty relations \cite{stringy-infl,stringy-infl1,stringy-infl2,stringy-infl3,stringy-infl4,stringy-infl5,stringy-infl6,stringy-infl7,stringy-infl8,stringy-infl9,stringy-infl10,stringy-infl11,stringy-infl12}. The power spectra of the scalar metric fluctuations in $\kappa$-deformed Robertson-Walker space time also contains non-trivial corrections \cite{kim-infl}. The inflationary models, constructed from the modified dispersion relations, due to the minimal length scenarios, have been shown to affect the propagation of fluctuation modes leading to modified Mukhanov-Sasaki equation and power spectra \cite{disp-infl1,disp-infl2,disp-infl3,disp-infl4,disp-infl5,disp-infl6,disp-infl7,disp-infl8,disp-infl9,disp-infl10,disp-infl11,disp-infl12,disp-infl13,disp-infl14,disp-infl15,disp-infl16,disp-infl17,disp-infl18,p}. Such modified Mukhanov-Sasaki equation has also been derived using $\kappa$-deformed dispersion relation \cite{kappa-disp-inf}. The primordial fluctuations arising from the deformed algebra between the inflaton field and its conjugate has also been shown to induce non-trivial modifications to the primordial power spectra \cite{deform1,deform2}.

Interestingly some of the non-commutative inflationary models have been examined in the light of observational data and constraints on the non-commutative parameters have also been placed using PLANCK \cite{nc-planck,nc-planck1}, WMAP \cite{nc-wmap,nc-wmap1,nc-wmap2}, ACBAR \cite{nc-acbar} and CBI \cite{nc-acbar} dataset. The ACT \cite{act1} has released data after investigating the inflationary scenario with recently acquired CMB data and a discrepancy with PLANCK's dataset \cite{planck} has been observed. This discrepancy could hints the possibility of new physics lurking in the very early universe and different works are progressing along this direction to provide a plausible explanation for this tension between different datasets \cite{tension1,tension2,tension3,tension4,tension5,tension6,tension7,tension8}. Recently ACT DR6 has has reported a higher value of scalar spectral index,  i.e., $n_s=0.974\pm 0.003$ \cite{act-ns}, in comparison to the PLANCK's value $n_s=0.965\pm 0.004$ \cite{planck}, and a positive running of spectral index $\alpha_s=0.0062 \pm 0.0052$ \cite{act-alpha}, different from $\alpha_s=-0.006 \pm 0.013$ of PLANCK \cite{planck}. In this work we suggest a possible resolution for these discrepancies within the framework of $\kappa$-deformed non-commutative space-time. 

We investigate the inflationary scenario in $\kappa$-deformed space-time where the space-time non-commutativity is introduced by the notion of $\kappa$-deformed oscillator algebra, which is derived by constructing a scalar field theory invariant under the $\kappa$-Poincaré algebra in the bi-crossproduct basis. This $\kappa$-deformed oscillator is shown to induce non-trivial modifications to the primordial power spectra and their spectral indices. The scale-dependent enhancement of the scalar spectral index due to the $\kappa$-deformation is seen to potentially resolve the ACT-PLANCK discrepancy, at extreme precision consistent with the minimal length scale description just a few orders of magnitude below the Planck length. Finally, we also perform a comprehensive Bayesian analysis using the latest ACT DR6 data \cite{act-ns,act-alpha} and places constraint on the non-commutative scale, which is in agreement with the value required to resolve the PLANCK-ACT discrepancy. 

This paper is organised in the following manner. In Sec.I, we construct the $\kappa$-deformed oscillator algebra for scalar field theory, whose equation of motion is obtained using the deformed Casimir operator of $\kappa$-Poincare algebra. In Sec.II, we calculate the modified power spectrum for scalar and tensor perturbations. Further we also calculate the deformed spectral indices and study the ACT-PLANCK discrepancy from it. In Sec.III, we perform a comprehensive Bayesian analysis and obtain an upper on the $\kappa$-deformation parameter. Finally in Sec.IV, we discuss our results and provide concluding remarks.  

%%%%%%%%%%%%%%%%%%%%%%%%%%%%%%%%%%%%%%%%%%%%%%%%%%%%%%%%%%%%%%%%%%%%%%%%%%%%%%%%%

\section{$\kappa$-deformed oscillator algebra}\label{sec1}

We begin with the discussion of $\kappa$-Poincare algebra and the deformed quadratic Caismir operator associated with it. Utilising this Casimir operator, we formulate the modified Klein-Gordon equation for the scalar field that remains invariant under the $\kappa$-deformed Poincare algebra. Subsequently, we derive the $\kappa$-deformed oscillator algebra for this by assuming a standard commutation relation between the $\kappa$-scalar field and its conjugate.

The $\kappa$-Poincare algebra, in the bi-crossproduct basis, is given as \cite{kpa2}
\begin{equation}\label{a1}
\begin{split}
 [M_i,M_j] &= \epsilon_{ijk} M_k,~~[N_i,N_j]=-\epsilon_{ijk}M_k,\\
 [M_i,N_j] &= \epsilon_{ijk}N_k,~~[P_i,P_0]=0,~~[P_i,P_j]=0,\\
 [M_i,P_0] &= 0,~~[M_i,P_j]=\epsilon_{ijk}P_k,~~[N_i,P_0]=-P_i,\\
 [N_i,P_j] &= \delta_{ij}\Big(\frac{1}{2\lambda}(1-e^{-2\lambda P_0}) + \frac{\lambda}{2}\vec{P}^2 \Big)-\lambda P_iP_j,
\end{split}
\end{equation}
where $\lambda=1/\kappa$ and $M_i=\frac{1}{2}\epsilon_{ijk}M_{jk}$ and $N_i=M_{i0}$ are the generators of rotation and boost, respectively. Note that in this particular basis, the Lorentz algebra sector remains undeformed. This $\kappa$-Poincare algebra contracts to the standard Poincare algebra in the limit $\lambda\to 0$. In this study, we use this limiting procedure to recover the commutative results. Now the invariant Casimir operator for this $\kappa$-Poincare algebra is obtained as \cite{kpa2}
\begin{equation}\label{a2}
 \mathcal{C}_{\lambda}^2=-\vec{P}^2e^{-\lambda P_0}+\Big(\frac{2}{\lambda}\sinh\frac{\lambda P_0}{2}\Big)^2.
\end{equation}
The Casimir operator obtained in Eq.(\ref{a2}), facilitates us to write down the scalar field equation in $\kappa$-Minkowski space-time as 
\begin{equation}\label{a3}
 \Big(\Box_{\lambda}+m^2\Big)\phi^{(\lambda)}(x)=0,
\end{equation} 
where $\Box_{\lambda}=-\partial_i^2e^{-i\lambda\partial_0}+\Big(\frac{2}{\lambda}\sin\frac{\lambda \partial_0}{2}\Big)^2$ is the deformed de-Alembertian, invariant under the $\kappa$-Poincare algebra. In genera the deformed field operator $\phi^{(\lambda)}(x)$ can be decomposed into Fourier modes as 
\begin{equation}\label{a2a}
 \phi^{(\lambda)}(x)=\displaystyle\int\frac{d^4p}{\sqrt{(2\pi)^4}} A(p)\delta(\mathcal{C}_{\lambda}^2-m^2)e^{-i(p_0t-\vec{p}\cdot\vec{x})}\Theta(p_0)
\end{equation} 
and it satisfies the equation of motion given in Eq.(\ref{a3}). Now we proceed our analysis by simplifying Eq.(\ref{a2a}) for the massless case, where we get $p_0=\frac{\ln(1+\lambda|\vec{p}|)}{\lambda}$ from the $\kappa$-dispersion relation. By simplifying $\delta(\mathcal{C}_{\lambda}^2)$ using $\delta(f(t))=\displaystyle \sum_i\frac{\delta(t-t_i)}{|f'(t_i)|}$ (where $f(t_i)=0$) and carrying out the resulting $p_0$ integral, we get $\phi^{(\lambda)}(x)$ as
\begin{equation}\label{a4}
\begin{split}
 \phi^{(\lambda)}(x)&=\int \frac{d^3\vec{p}}{\sqrt{2(2\pi)^3|\vec{p}|^2}}\bigg(\hat{a}(\vec{p}) e^{-i\big(\frac{\ln(1+\lambda|\vec{p}|)t}{\lambda}-\vec{p}\cdot\vec{x}\big)} + \\  &~~~\hat{a}^{\dagger}(\vec{p}) e^{i\big(\frac{\ln(1+\lambda|\vec{p}|)t}{\lambda}-\vec{p}\cdot\vec{x}\big)}\bigg),
\end{split}
\end{equation}
where $\hat{a}^{\dagger}(\vec{p})$ and $\hat{a}(\vec{p})$ are the $\kappa$-deformed creation and annihilation operators, respectively.

The conjugate momenta corresponding to this field operator is defined as $\pi^{(\lambda)}(x)=\partial_0\phi^{(\lambda)}(x)$. Using Eq.(\ref{a4}), we calculate the commutation relation between this deformed field operator and its conjugate. This commutation relation retains the standard form, i.e., $[\phi^{(\lambda)}(\vec{x},t),\pi^{(\lambda)}(\vec{x}',t)]=i\delta(\vec{x}-\vec{x}')$, by assuming the following deformed oscillator algebra
\begin{equation}\label{a5}
\begin{split}
 [\hat{a}(\vec{p}),\hat{a}^{\dagger}(\vec{p}')]&=\frac{\lambda|\vec{p}|}{\ln(1+\lambda|\vec{p}|)}\delta^3(\vec{p}-\vec{p}'),\\
 [\hat{a}(\vec{p}),\hat{a}(\vec{p}')]&=0,~~[\hat{a}^{\dagger}(\vec{p}),\hat{a}^{\dagger}(\vec{p}')]=0.
\end{split}
\end{equation}
The deformation factor appearing in Eq.(\ref{a5}) depends on the $\kappa$-dispersion relation given in Eq.(\ref{a2}). So the form of this deformation factor changes when the $\kappa$-deformed dispersion relation is realised in other basis. However in the commutative limit all these expression reduce to the standard ones consistently. Note that the $\kappa$-deformed oscillator algebra obtained in Eq.(\ref{a5}) serves as the foundation for calculating the primordial power spectrum to study inflationary paradigm in this non-commutative setting.

%%%%%%%%%%%%%%%%%%%%%%%%%%%%%%%%%%%%%%%%%%%%%%%%%%%%%%%%%%%%%%%%%%%%%%%%%%%%%%%%%%%%%%%%%%%%%%%%%%%%%%%%%%%%%%%%%%%%%%%%%%%%%%%%%

\section{Primordial power spectra in $\kappa$-deformed space-time}

We consider a spatially flat FLRW space-time background endowed with linearised scalar and tensor perturbations. These tensor and scalar pertubations modes can be decoupled in accordance with the decomposition theorem and hence we can study their evolutions independently. Thus the evolution of the tensor perturbation mode is given as \cite{dodelson}
\begin{equation}\label{b1} 
 h_T'' + 2\frac{a'}{a}h_T' + k^2 h_T =0.
\end{equation}
Here $h_T$ represents the two polarisation components $(h_+,~h_{\times})$ of the tensor perturbation $h_{ij}$, $k$ is the comoving wavenumber (whereas $p$ in Eq.(\ref{a5}) is the physical momenta \cite{p}) and $'$ denotes a derivative with respect to conformal time $\tau$. We define ${h}=\frac{ah_T}{\sqrt{16\pi G}}$ and decompose the field operator corresponding to the tensor perturbation using creation and annihilation operators as
\begin{equation}\label{b1a}
 \hat{{h}}({k},\tau)=\nu({k},\tau)\hat{a}({k})+\nu^*({k},\tau)\hat{a}^{\dagger}({k}).
\end{equation}
Substituting the above defined field decomposition of tensor perturbation, i.e. Eq.(\ref{b1a}), in Eq.(\ref{b1}), we get the Mukhanov-Saski equation as ${\nu}'' + \Big(k^2 -\frac{a''}{a}\Big){\nu}=0$. By considering the slow-roll inflation and using the Bunch-Davies initial condition $\displaystyle\nu(k,\tau)=\frac{e^{-ik\tau}}{\sqrt{2k}}$, the  solution is obtained as \cite{dodelson}
\begin{equation}\label{b1b}
 \nu(k,\tau)=\displaystyle\frac{e^{-ik\tau}}{\sqrt{2k}}\Big(1-\frac{i}{k\tau}\Big)
\end{equation}
In this analysis we incorporate the effects of $\kappa$-deformed space-time non-commutativity in the inflationary paradigm through the $\kappa$-deformed oscillator algebra obtained in Eq.(\ref{a5}). Therefore we assume the creation and annihilation operators to obey the following $\kappa$-deformed oscillator algebra as
\begin{equation}\label{b3}
\begin{split}
 [\hat{a}({k}),\hat{a}^{\dagger}({k}')]&=\frac{\lambda{k}/a}{\ln(1+\lambda{k}/a)}\delta^3(k-{k}'),\\
 [\hat{a}({k}),\hat{a}({k}')]&=0,~~[\hat{a}^{\dagger}({k}),\hat{a}^{\dagger}({k}')]=0
\end{split}
\end{equation}
The standard power spectrum for $h$ is defined as $\mathcal{P}_h(k)=\langle\hat{h}^{\dagger}_T({k},\tau) \hat{h}_T({k}',\tau)\rangle$ \cite{dodelson}. Using the field decomposition given in Eq.(\ref{b1a}) and the $\kappa$-deformed oscillator algebra given in Eq.(\ref{b3}), we get the deformed power spectrum for $h$ as $\displaystyle\mathcal{P}^{(\lambda)}_h(k)=\mathcal{P}^{(0)}_h(k) \frac{\lambda{k}/a}{\ln(1+\lambda{k}/a)}$, where $\mathcal{P}^{(0)}_h(k)=\displaystyle 16\pi G\frac{|\nu(k,\tau)|^2}{a^2}$. This mode exit the horizon and becomes $\nu(k,\tau)\simeq\displaystyle-\frac{i}{k\tau}\frac{e^{-ik\tau}}{\sqrt{2k}}$ when $k|\tau| \ll 1$. As a result the amplitude of tensor perturbation freezes and remains constant thereafter. Now we consider the slow roll inflation where the Hubble rate vary slowly. Thus under this slow-roll approximation, we get $\tau\simeq-1/aH$ and at the horizon crossing it becomes $\tau\simeq-1/k$. From this we find $\mathcal{P}^{(0)}_h(k)=\displaystyle\frac{8\pi GH^2}{k^3}$ \cite{dodelson}. By using the definition $\mathcal{P}^{(\lambda)}_T=k^3\mathcal{P}^{(\lambda)}_h(k)$, we obtain the $\kappa$-deformed tensor power spectrum at the moment of mode exiting the horizon as
\begin{equation}\label{b5}
 \mathcal{P}^{(\lambda)}_T=\mathcal{P}^{(0)}_T \frac{\lambda H}{\ln(1+\lambda H)}\bigg|_{k=aH}
\end{equation}
where $\mathcal{P}^{(0)}_T=8\pi G H^2$ is the standard tensor power spectrum.

Now we look upon the evolution  of the quantum fluctuations of the inflaton field $\phi$. The equations of motion for the first order perturbation of this scalar field is given by \cite{dodelson} 
\begin{equation}\label{b6}
 \delta\phi'' + 2\frac{a'}{a}\delta\phi' +  (k^2 +a^2 V_{,\phi\phi}) \delta\phi=0,
\end{equation}
where $V$ is the potential of scalar field and $V_{,\phi\phi} =d^2V/d\phi^2$. 
During the slow-roll inflation, %$V_{,\phi\phi}<<H^2$ and hence
  Eq.(\ref{b6}) can be taken as $\delta\phi'' + 2\frac{a'}{a}\delta\phi' + k^2 \delta\phi\simeq 0$ under the sub-horizon limit \cite{dodelson}. This expression is similar to the evolution of the tensor perturbation equation written in Eq.(\ref{b1}) and hence  we can use the same procedure to calculate the power spectrum associated with the fluctuations of inflaton  field. Thus we decompose the fluctuation $\delta\phi$ in terms of creation and annihilation operators, satisfying the $\kappa$-deformed oscillator algebra as in Eq.(\ref{b3}). By employing this deformed oscillator algebra we calculate $\langle\delta\phi(k,\tau)\delta\phi(k',\tau)\rangle$ using the above outlined steps for tensor perturbation and get the deformed power spectrum for $\delta\phi$ as $\mathcal{P}^{(\lambda)}_{\delta\phi}(k)=\displaystyle\mathcal{P}^{(0)}_{\delta\phi}(k) \frac{\lambda{k}/a}{\ln(1+\lambda{k}/a)}$, where $\mathcal{P}^{(0)}_{\delta\phi}(k)=\displaystyle \frac{H^2}{2k^3}$ \cite{dodelson}.

The metric perturbations become significant in the post-inflationary regime as the perturbation from the scalar field inflaton gets transferred to the scalar part of the metric perturbation by the end of inflation at horizon crossing. This scalar part of the metric perturbation $\Phi$ can be expressed in terms of the fluctuations of inflaton at horizon crossing and as a result the deformed power spectrum corresponding to $\Phi$ becomes $\mathcal{P}^{(\lambda)}_{\Phi}(k)=\displaystyle\frac{16\pi G}{9\varepsilon}\mathcal{P}^{(\lambda)}_{\delta\phi}(k)\Big|_{k=aH}$, where $\varepsilon=-H'/H^2$ is the slow-roll parameter \cite{dodelson}. Now using the definition $\mathcal{P}^{(\lambda)}_s=\displaystyle{k^3}\mathcal{P}^{(\lambda)}_{\Phi}(k)$ , we obtain the $\kappa$-deformed scalar power spectrum as
\begin{equation}\label{b7}
 \mathcal{P}^{(\lambda)}_s=\mathcal{P}^{(0)}_s \frac{\lambda H}{\ln(1+\lambda H)}\bigg|_{k=aH}
\end{equation}
where $\mathcal{P}^{(0)}_s=\displaystyle\frac{8\pi G H^2}{9\varepsilon}$ is the standard scalar power spectrum. It is interesting to see from Eq.(\ref{b5}) and Eq.(\ref{b7}) that the scalar-to-tensor ratio $r$ remains unaffected under the above prescription of $\kappa$-deformation.

The power spectra can be conventionally written as $\mathcal{P}_T\equiv A_T k^{n_T}$ and $\mathcal{P}_s\equiv A_s k^{n_s-1}$, where $A_T$ and $A_s$ are the tensor and scalar amplitudes \cite{dodelson}. From these relations it is straightforward to get the expression for scalar and tensor spectral indices as $n_s^{(\lambda)}-1=\displaystyle\frac{d\ln\mathcal{P}^{(\lambda)}_s}{d\ln k}$ and $n_T^{(\lambda)}=\displaystyle\frac{d\ln\mathcal{P}^{(\lambda)}_T}{d\ln k}$, respectively. By using Eq.(\ref{b5}) and $\displaystyle\frac{d\ln H}{d\ln k}=-\varepsilon$ \cite{dodelson}, we find explicit form of $\kappa$-deformed tensor spectral index as  
\begin{equation}\label{b8}
 n_T^{(\lambda)}=n_T^{(0)} - \varepsilon + \frac{\varepsilon\lambda H}{\ln(1+\lambda H)}
\end{equation}
where $n_T^{(0)}=-2\varepsilon$ is the standard tensor spectral index. Similarly by substituting Eq.(\ref{b6}) in the above definition and using  $\displaystyle\frac{d\ln\varepsilon}{d\ln k}=2(\varepsilon+\delta)$ \cite{dodelson}, we find the $\kappa$-deformed scalar spectral index as 
\begin{equation}\label{b9}
 n_s^{(\lambda)}= n_s^{(0)} - \varepsilon + \frac{\varepsilon\lambda H}{\ln(1+\lambda H)}
\end{equation}
where $n_s^{(0)}=1-6\varepsilon+2\eta$ is the standard scalar spectral index and $\eta=\frac{1}{8\pi G}\frac{V_{,\phi\phi}}{V}$ is the second slow roll parameter \cite{dodelson}. Recently the ACT-DR6 \cite{act-ns} has reported a higher value for the scalar spectral index, $n_s = 0.974 \pm 0.003$ as compared to ones reported by PLANCK $n_s=0.965\pm 0.004$. We expect the minimal length scale correction associated with the space-time non-commutativity could a provide plausible explanation for this discrepancy. For this we consider the $n_s^{(\lambda)}$ as the scalar spectral index of ACT and $n_s^{(0)}$ as that of PLANCK. In order to analyse this we take the expression for $n_s^{(\lambda)}$, given in Eq.(\ref{b9}), valid up to first order in $\lambda$, as 
\begin{equation}\label{b9a}
 n_s^{(\lambda)}= n_s^{(0)} + {\lambda H\varepsilon}/{2}
\end{equation}
Note that the above expression has been obtained by setting $\lambda$ to be negative. In general $\lambda$ can be either positive or negative as it is the non-commutative parameter associated with $\kappa$-deformed space-time whose magnitude corresponds to the physically observable length scale and hence it is legitimate to fix $\lambda$ to be negative in this study. Comparing the difference between the scalar spectral indices of ACT and PLANCK with Eq.(\ref{b9a}), we get $\lambda\varepsilon H/2\simeq 0.009$. By setting $\varepsilon\simeq 0.007$ and $H\simeq 10^{14}$GeV, we obtain $\lambda\leq 5.07\times 10^{-30}m$. 

The running of the scalar spectral index in $\kappa$-deformed space-time can be calculated from $\alpha^{(\lambda)}_s=\displaystyle\frac{d n_s^{(\lambda)}}{d\ln k}$ using Eq.(\ref{b9})
\begin{equation}\label{b10}
\begin{split}
 \alpha^{(\lambda)}_s = & \alpha^{(0)}_s - 2\varepsilon(2\varepsilon-\eta) + \frac{2\lambda H\varepsilon(2\varepsilon-\eta)}{(1+\lambda H)\ln(1+\lambda H)} \\
 &- \frac{\lambda H\varepsilon^2}{(1+\lambda H)\ln(1+\lambda H)} + \frac{\lambda^2 H^2\varepsilon^2}{(1+\lambda H)^2\ln(1+\lambda H)}\\
 &+ \frac{\lambda^2 H^2\varepsilon^2}{(1+\lambda H)^2(\ln(1+\lambda H))^2}
\end{split}
\end{equation}
where $\alpha^{(0)}_s = -24\varepsilon^2 + 16\varepsilon\eta - 2\zeta$ is the standard running of scalar spectral index and $\zeta=\frac{1}{64\pi^2 G^2}\frac{V_{,\phi}V_{,\phi\phi}}{V^2}$ is the third slow roll parameter. By following the arguments as in Eq.(\ref{b9a}), we get the running of spectral index valid up to $\lambda$ term as
\begin{equation}\label{b11}
 \alpha_s^{(\lambda)} =-24\varepsilon^2 + 16\varepsilon\eta - 2\zeta - \lambda H\varepsilon\eta + 3{\lambda H\varepsilon^2}/{2}
\end{equation}
where the first order correction to $\kappa$-deformation induces two correction terms in the running of spectral index when compared to one correction term appearing in the expression for the spectral index (as seen in Eq.(\ref{b9a})).

%%%%%%%%%%%%%%%%%%%%%%%%%%%%%%%%%%%%%%%%%%%%%%%%%%%%%%%%%%%%%%%%%%%%%%%%%%%%%%%%%%%%%%%%%%%%%%%%%%%%%%%%%%%%%%%%%
\section{Observational constraints}

We present a comprehensive Bayesian analysis constraining the minimal length scale $\lambda$, associated with the $\kappa$-deformed space-time non-commuativity. For this we utilise the expressions for the $\kappa$-deformed scalar spectral index and its running derived in Eq.(\ref{b9a}) and Eq.(\ref{b11}). Using the Markov Chain Monte Carlo (MCMC) approach with the latest ACT DR6 data \cite{act-ns,act-alpha}, we obtain constraints on the non-commutative parameter $\lambda$ and the slow roll parameters.

We implement this MCMC sampling using the \textit{emcee} Python package. In this analysis we consider the following likelihood function
\begin{equation}\label{c1}
 \ln\mathcal{L}=-\frac{1}{2}\bigg[ \Big(\frac{n_s^{th}(\theta)-n_s^{obs}}{\sigma_{n_s}}\Big)^2 + \Big(\frac{\alpha_s^{th}(\theta)-\alpha_s^{obs}}{\sigma_{\alpha_s}}\Big)^2 \bigg]
\end{equation}
where $\theta=(\varepsilon,\eta,\zeta,\lambda)$ are the model parameters. Here $n_s^{th}(\theta)$ and $\alpha_s^{th}(\theta)$ are given by Eq.(\ref{b9a}) and Eq.(\ref{b11}). From \cite{act-ns,act-alpha} we set $n_s^{obs}=0.974\pm 0.003$ and $\alpha_s^{obs}=  0.0062 \pm 0.0052$. The slow-roll parameters $\varepsilon,\eta,\zeta$ are imposed with Gaussian priors values $[0.006,0.008],~[0.007,0.011],~[0.0007,0.0017]$, respectively. For $\lambda$ we set a uniform log prior as $[10^{-34},10^{-26}]$

%We implement the MCMC sampling using the emcee Python package []. It employs 100 Markov chains for 10,000 iterations, discarding the initial 20\% as convergence phase and applying a subsampling interval of 50 generating approximately 16,000 independent samples with an effective sample size exceeding 1000 for all parameters, ensuring reliable posterior inference. 
\begin{widetext} 
\begin{center}
\begin{figure}
\includegraphics[width=0.75\textwidth]{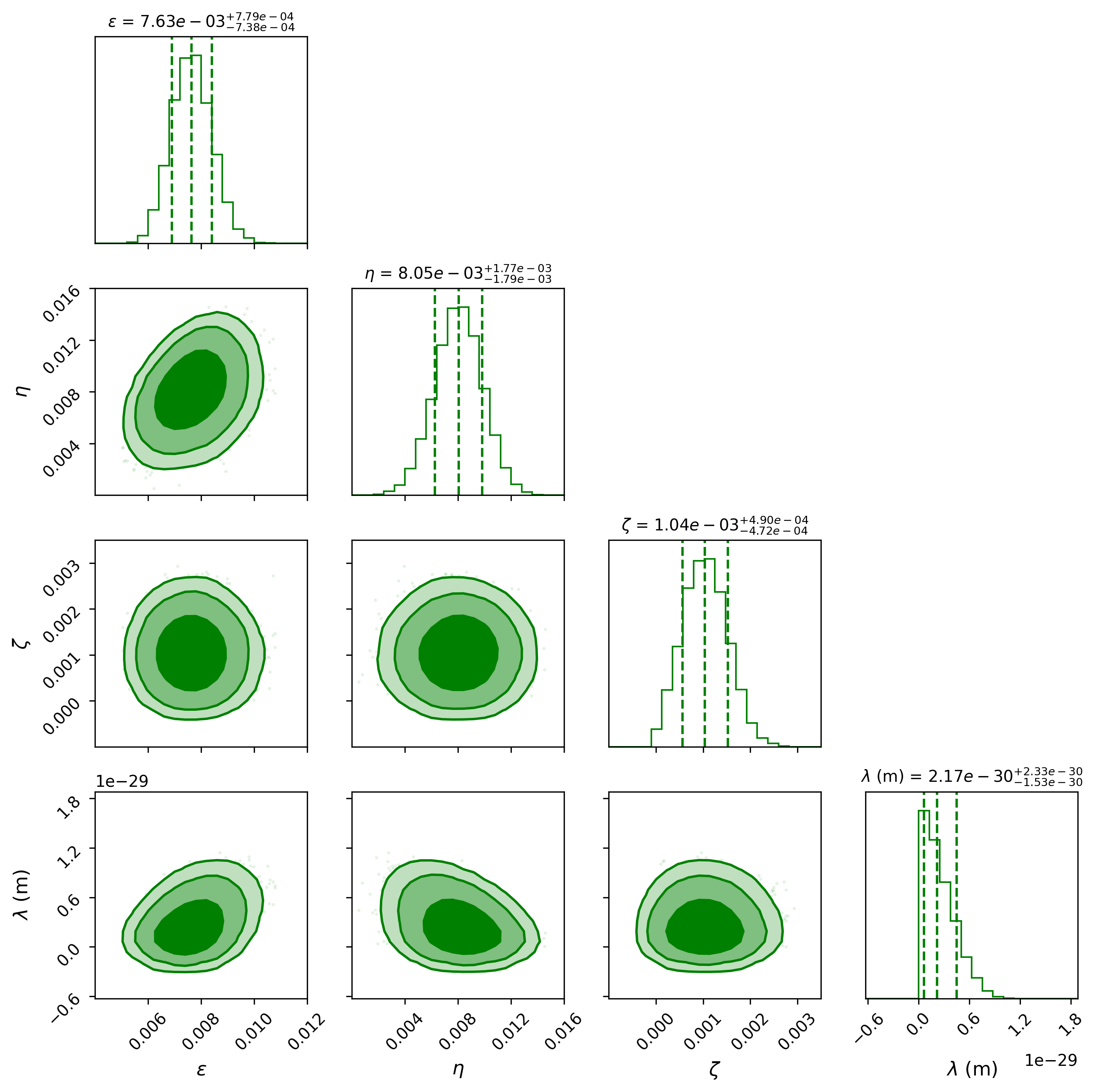}
\caption{The marginalised posterior distribution of the slow-roll parameters and $\kappa$-deformation scale derived using the latest ACT DR6 measurements of scalar spectral index and its running} \label{fig:corner}
\end{figure}
\end{center}

\begin{center}  
\begin{table}
\centering  
\begin{tabular}{c c c c c c}
\multicolumn{6}{c}{ }\\
\hline 
   &&&&& \\
  Parameter~~~~~~~&~~~~~~~Marginalised value~~~~~~~&~~~~~~~$1\sigma$ Credible Interval~~~~~~~&~~~~~~~$2\sigma$ Credible Interval   \\
 \hline 
   
    &&& \\
   $\lambda$ & $2.17\times 10^{-30}$ & $[0.64,4.50]\times 10^{-30}$ & $[0.10,7.08]\times 10^{-30}$\\
   $\varepsilon$ & $7.63\times 10^{-3}$ & $[6.85,8.41]\times 10^{-3}$ & $[6.21,9.17]\times 10^{-3}$\\
   $\eta$ & $8.05\times 10^{-3}$ & $[6.26,9.82]\times 10^{-3}$ & $[0.44,1.15]\times 10^{-2}$\\
   $\zeta$ & $1.04\times 10^{-3}$ & $[0.57,1.53]\times 10^{-3}$ & $[0.20,2.00]\times 10^{-3}$\\
 \hhline{------}   
\end{tabular} 
\caption{\small {The marginalized values with $1\sigma$ and $2\sigma$ credible intervals of slow-roll parameters and $\kappa$-deformed length scale}}\label{tab:const}
\end{table} 
\end{center}
\end{widetext}

The inner, middle and outer regions of Fig.(\ref{fig:corner}) show the $1\sigma,~2\sigma$ and $3\sigma$ confidence level contour plot of the $\kappa$-deformed and slow roll parameters, respectively. Based on this analysis depicted in Fig.(\ref{fig:corner}) we obtain well constrained slow-roll parameters having nearly Gaussian posterior distribution, whose marginalized values including $1\sigma$ and $2\sigma$ credible intervals are shown in Table.(\ref{tab:const}). We observe an asymmetric posterior distribution for the $\kappa$-deformed length scale, likely due to the physical requirement $\lambda>0$. The $\lambda$ is constrained to a marginalized value at $1\sigma$ confidence level as $2.17^{+2.33}_{-1.53}\times 10^{-30}m$ within a $2\sigma$ of $6.30\times 10^{-30}m$. This $\lambda$ obtained from MCMC analysis is in close agreement with the value $5.07\times 10^{-30}m$ required for the potential resolution of ACT-PLANCK discrepancy over spectral index. %By substituting the marginalised values of Table.(\ref{tab:const}) in Eq.(\ref{b9a}) and Eq.(\ref{b11}), we get $n_s=0.974$ and $\alpha_s=-0.0025$. This value of spectral index obtained from the above MCMC analysis is in exact agreement with ACT's value $n_s=0.974\pm 0.003$ \cite{act-ns}. Although the obtained running spectral value is negative it is consistent with the ACT's result  $\alpha_s=0.0062 \pm 0.0052$  at $2\sigma$ confidence level \cite{act-alpha}.

%%%%%%%%%%%%%%%%%%%%%%%%%%%%%%%%%%%%%%%%%%%%%%%%%%%%%%%%%%%%%%%%%%%%%%%%%%%%%%%%%%%%%%%%%%%%%%%%%%%%%%%%%%%%%%%%%
\section{Conclusion}

We have investigated the imprints of space-time non-commutativity on the inflationary paradigm within the framework $\kappa$-deformed non-commutative space-time using the ACT DR6 dataset. By utilisling the deformed Casimir operator of the $\kappa$-Poincare algebra we have constructed a scalar field theory (invariant under the $\kappa$-Poincaré algebra) and subsequently derived the deformed oscillator algebra in $\kappa$-deformed space-time, where the form of this deformation factor strongly depends on the choosen basis of the $\kappa$-Poincare algebra.  This $\kappa$-deformed oscillator algebra plays a crucial role in incorporating the non-commutative effects into the dynamics of primordial perturbations.

The power spectra of scalar as well as tensor perturbations in the $\kappa$-deformed space-time non-commutative space-time obtained using the $\kappa$-deformed oscillator algebra under the standard slow-roll inflationary regime are shown to have a non-trivial scale dependent modifications. These modifications are expected to encode the possible quantum gravity signatures coming from the early universe. Furthermore the deformed oscillator algebra also induces scale dependent corrections to the spectral indices which some has some observational consequences. Such non-trivial scale dependency has also been shown to appear while studying the inflation in other non-commutative space-time models \cite{moyal-infl,moyal-infl1,moyal-infl2,moyal-infl3,moyal-infl4,moyal-infl5,moyal-infl6,moyal-infl7,moyal-infl8}. Although the $\kappa$-deformation modifies the amplitudes of both the scalar and tensor power spectra, the scalar-to-tensor ratio remains intact as the modification to the primordial power spectra is governed solely by the deformed oscillator algebra.

The $\kappa$-deformed space-time non-commutativity framework provides a potential resolution to the discrepancy between the values of the scalar spectral index  reported by the PLANCK and ACT collaborations. By identifying the standard spectral index with PLANCK's value and the $\kappa$-deformed spectral index with ACT's value, we obtained a bound for the non-commutative length scale as $\lambda\leq 5.07\times 10^{-30}m$.

We have implemented a comprehensive Bayesian Markov Chain Monte Carlo (MCMC) analysis using the latest ACT DR6 data yielding a marginalised value for the minimal length scale as $\lambda=2.17^{+2.33}_{-1.53}\times 10^{-30}m$ at the $1\sigma$ confidence level with a $2\sigma$ upper limit of $6.30\times 10^{-30}m$. This result corroborates the above vaule and provides a numerical validation for our analysis. Our constraint scale lies about four orders of magnitude below the Planck length and is a consistent description of minimal length scale in the context of space-time non-commutativity within the realm of quantum gravity phenomenology suggesting that the high precision cosmology can probe these extreme fundamental scales.

Eventhough the direct detection of $\kappa$-deformed space-time non-commutativity using the current observational capabilities remains obscure due to weak constraints and the degeneracies between scale-dependent modification of the scalar spectral index and standard inflationary parameters, the observed higher value of spectral index in ACT DR6 in comparison to PLANCK provides an intriguing consistency with the predicted $\kappa$ deformation-induced spectral index. Further improvements can be achieved by adopting self-consistent treatment where one should derive the modified inflationary dynamics directly from the $\kappa$-deformed gravitational action and such a study could potentially achieve a conclusive evidence for this quantum-gravitational signature in the primordial universe.

\section*{Acknowledgment}

This work is supported by the National Natural Science Foundation of China (Grant No.~12275080), and the Innovative Research Group of Hunan Province (Grant No.~2024JJ1006).

%%%%%%%%%%%%%%%%%%%%%%%%%%%%%%%%%%%%%%%%%%%%%%%%%%%%%%%%%%%%%%%%%%%%%%%%%%%%%%%%%%%%%%%%%%%%%%%%%%%%%%%%%%%%%%%%%

%\section*{}

\end{document}